\documentclass[letterpaper, 10 pt, conference]{ieeeconf}  

\IEEEoverridecommandlockouts                              

\usepackage{graphicx}
\usepackage{url}  

\usepackage{amsmath}

\usepackage{wrapfig}
\usepackage{mathtools}

\title{\LARGE \bf
Deep Learning Based Multimodal with Two-phase Training Strategy for Daily Life Video Classification
}

\author{Lam~Pham$^{1*}$, 
        Trang~Le$^{2*}$,
        Cam~Le$^{3}$, 
        Dat~Ngo$^{4}$, 
        Weißenfeld~Axel$^{5}$, 
        Alexander~Schindler$^{6}$ %
\thanks{L. Pham, W. Axel, and A. Schindler are with Center for Digital Safety \& Security, Austrian Institute of Technology, Austria.}%
\thanks{T. Le is with JVN Institute, VNU, Vietnam.}%
\thanks{C. Le is with HCM University of Technology, Vietnam.}%
\thanks{D. Ngo is with School of Computer Science and Electronic Engineering, University of Essex, UK.}%
\thanks{(*) Main and equal contribution into the paper.}
}

\begin{document}

\maketitle
\thispagestyle{empty}
\pagestyle{empty}

\begin{abstract}
In this paper, we present a deep learning based multimodal system for classifying daily life videos.
To train the system, we propose a two-phase training strategy. 
In the first training phase (Phase I), we extract the audio and visual (image) data from the original video.
We then train the audio data and the visual data with independent deep learning based models.
After the training processes, we obtain audio embeddings and visual embeddings by extracting feature maps from the pre-trained deep learning models. 
In the second training phase (Phase II), we train a fusion layer to combine the audio/visual embeddings and a dense layer to classify the combined embedding into target daily scenes.
Our extensive experiments, which were conducted on the benchmark dataset of DCASE (IEEE AASP Challenge on Detection and Classification of Acoustic Scenes and Events) 2021 Task 1B Development, achieved the best classification accuracy of 80.5\%, 91.8\%, and 95.3\% with only audio data, with only visual data, both audio and visual data, respectively. 
The highest classification accuracy of 95.3\% presents an improvement of 17.9\% compared with DCASE baseline and shows very competitive to the state-of-the-art systems.
\end{abstract}

\section{INTRODUCTION}
\label{intro}
Recently, applying deep learning techniques to analyze videos has achieved many successes and opened a variety of real-life applications. 
Indeed, a wide range of deep learning based systems have been proposed for various human-relevant tasks of emotion recognition~\cite{cite_06}, lip-reading~\cite{cite_02}, or detecting human activities~\cite{cite_10, sui2020real, cite_04}, etc.
Recently, a dataset of daily-scene videos~\cite{ds_2021_1b}, which was proposed by DCASE challenge~\cite{dcase_web} for a new task of audio-visual scene classification (AVSC), was published and attracted attention from the video research community.
Similar to the systems proposed for analyzing videos of human activities~\cite{cite_04, cite_06}, the state-of-the-art systems proposed for AVSC task also leveraged deep learning based models and presented joined audio-visual analysis.
For instances, the proposed systems in~\cite{pp_av_01, pp_av_02} used convolutional based models to extract audio embeddings from audio data and leveraged pre-trained deep learning models for extracting visual embeddings from visual data.
Then, the audio embeddings and the visual embeddings are concatenated and fed into dense layers for classification.
To further enhance audio/visual embeddings, the authors in~\cite{pp_av_03} proposed a graphed based model which was used to learn the audio/visual feature maps extracted from middle layers of deep learning backbone models.
The graph based model then generates a graph based audio/visual embedding.
The graph based audio/visual embeddings are finally fused with audio/visual embeddings before going to dense layers for classification.
Meanwhile, authors in~\cite{pp_av_04} improved the audio/visual embeddings by proposing a contrastive event-object alignment layer.
The contrastive event-object alignment layer, which is based on the contrastive learning technique, helps to explore the relationship between audio and visual information by learning relative distances of event-object pairs occurring in both audio and visual scenes.

In this paper, we also leverage deep learning techniques, propose a deep learning based multimodal system for the task of AVSC. 
We present our main contributions:
(a) We propose a mechanism, which combines a multi-model fusion and a two-phase training strategy, to generate an audio-visual embedding representing for one video input.
(b) We evaluate our proposed deep learning based multimodal system on the DCASE 2021 Task 1B Development set which is the benchmark and largest dataset for the task of audio-visual daily scene classification. 
Results reveal that our proposed system is very competitive to the state-of-the-art systems.

\section{Proposed Deep Learning Based Multimodal for AVSC task}
\label{frameworks}
\begin{figure}[t]
    \centering
    \includegraphics[width =1.0\linewidth]{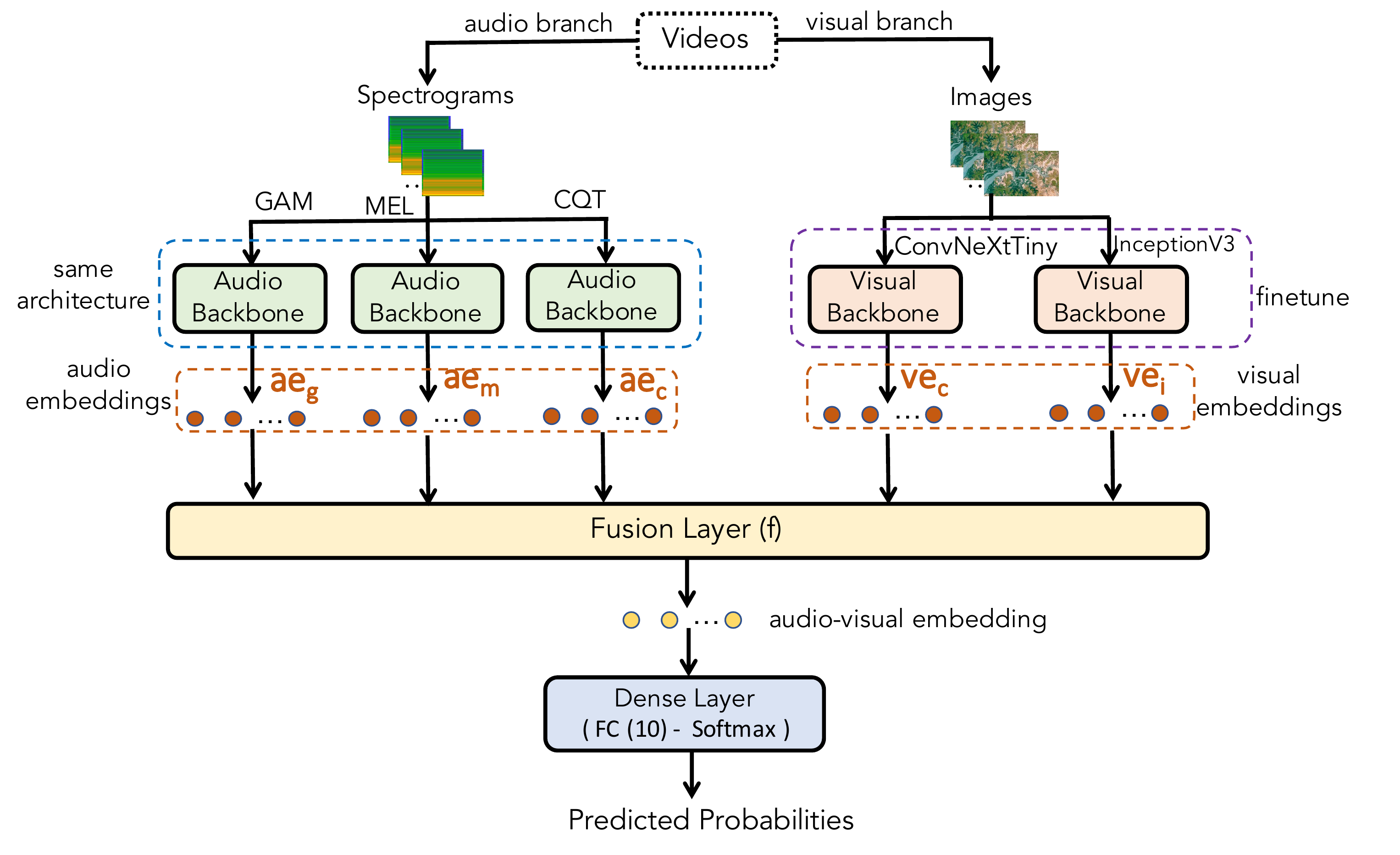}
    	\vspace{-0.5cm}
	\caption{The high-level architecture of the proposed deep learning based multimodal for AVSC task}
    \label{fig:f1}
\end{figure}
As Figure~\ref{fig:f1} shows, the high-level architecture of our proposed deep learning based multimodal for audio-visual scene classification (AVSC) comprises two individual branches: the audio branch and the visual branch, which focus on either audio or visual data extracted from the input video. 
Regarding the audio branch, the audio is first transformed into spectrograms which are then fed into three Audio Backbones to extract audio embeddings.
Meanwhile, images are fed into two Visual Backbones to extract image embeddings.
The audio and image embeddings are then combined by a Fusion Layer to generate an audio-visual embedding (i.e. The Fusion Layer is denoted by the function $f$).
The audio-visual embedding is finally classified into target categories by a Dense Layer.
From results shown in recent papers~\cite{pp_av_01, pp_av_02, pp_av_04}, we can see that the visual data contributes to the AVSC performance more significantly than the audio data.
If we train our proposed AVSC system with an end-to-end training process, it possibly causes an overfitting on the visual branch and reduces the role of the audio branch.  
We therefore propose a two-phase training strategy to train our proposed AVSC system.
While the first training phase (Phase I) is used to train and achieve the individual Audio and Visual Backbones, the Fusion Layer and the Dense Layer are trained in the second phase (Phase II).

\begin{figure}[t]
    \centering
    \includegraphics[width =1.0\linewidth]{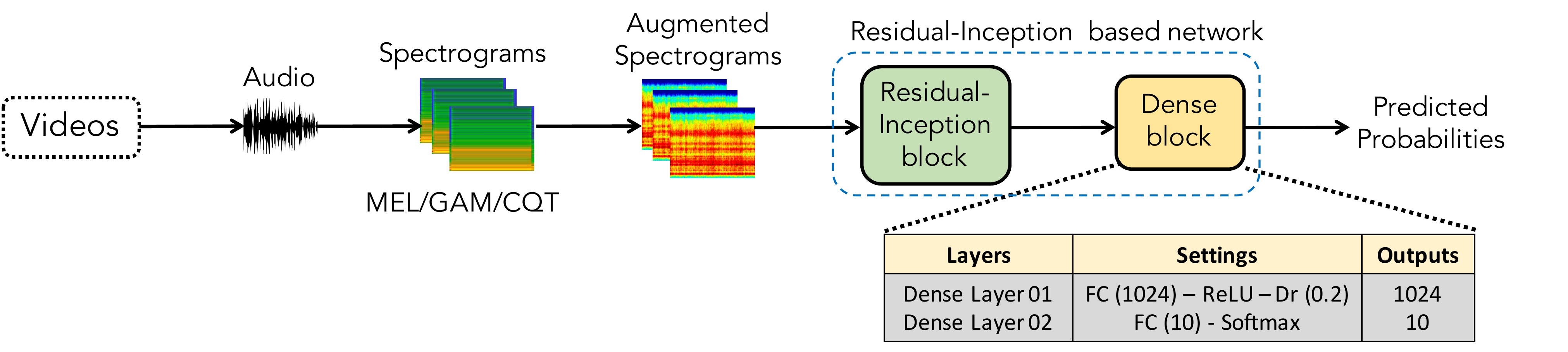}
    	\vspace{-0.5cm}
	\caption{Deep learning based models with audio data}
    \label{fig:f2}
\end{figure}
\subsection{Phase I: Train deep learning models on individual audio or visual data to achieve audio and visual backbones}
\label{phase1}

In Phase I, we aim to achieve individual high-performance Audio and Visual Backbones as shown in Figure~\ref{fig:f1}.
To this end, we consider the AVSC task as a combination of two independent tasks of Acoustic Scene Classification (ASC) and Visual Scene Classification (VSC).
To deal with the ASC task, we leverage multiple input spectrograms, which proves powerful to improve the ASC performance~\cite{lam12,lam02}.
In particular, we propose deep learning based systems as shown in Figure~\ref{fig:f2} to train audio data.
The audio is firstly re-sampled to 32,000 Hz, then transformed into three types of spectrogram: Mel spectrogram (MEL), Gammatone (GAM), and Constant-Q-Transform (CQT) where both temporal and spectral features are presented. 
By using two channels and setting parameters of the filter number, the window size, the hop size to 128, 80 ms, 14 ms, respectively, we generate MEL, GAM, and CQT spectrograms of $128{\times}309{\times}2$ from one 10-second audio segment.
Delta and delta-delta are then applied to the three-dimensional spectrograms to obtain six-dimensional spectrograms of $128{\times}305{\times}6$. 
Next, the Mixup~\cite{mixup2} data augmentation method is applied on the six-dimensional spectrograms before feeding into a residual-inception based network for classification.
Regarding the residual-inception based network for training audio spectrograms, it is separated into two main parts: A Residual-Inception block and a Dense block.
The Residual-Inception block in this paper is the CNN-based backbone of the novel residual-inception deep neural network architecture which is reused from our previous works in~\cite{pham2022robust}.
Meanwhile, the Dense block comprises two dense layers with detailed configuration shown in the lower part of Figure~\ref{fig:f2}.
As we apply three types of spectrogram transformation (e.g. MEL, GAM, and CQT), we obtain three individual deep learning based models for audio input, referred to as the Aud-MEL, Aud-GAM, Aud-CQT models, respectively.
\begin{figure}[t]
    \centering
    \includegraphics[width =1.0\linewidth]{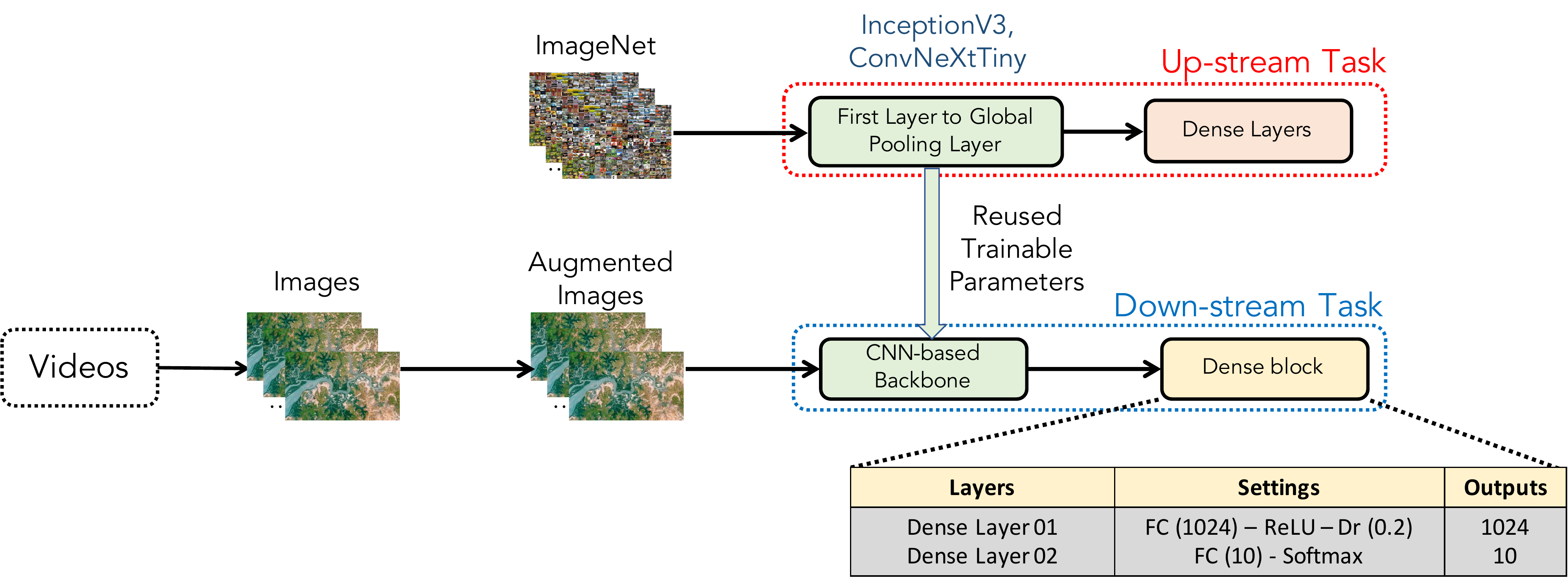}
    	\vspace{-0.5cm}
	\caption{Deep learning based models with visual data}
    \label{fig:f3}
\end{figure}

As regards the visual data, we propose deep learning models as shown in Figure~\ref{fig:f3}.
As Fig~\ref{fig:f3} shows, the original images (i.e. two images from each second) extracted from scene videos are first scaled into the tensor of $224{\times}224{\times}3$ with RGB format.
Then, the Mixup~\cite{mixup2} data augmentation method is applied on the scaled images before feeding into classification models.
To construct the classification models, we are inspired by~\cite{yu2023inceptionnext} which shows that a combination of Inception based and ConvNet based models proves effective to improve the performance of VSC tasks.
We, therefore, select InceptionV3 and ConvNeXtTiny networks from Keras library~\cite{keras_app} for evaluating the VSC task in this paper.
As both  InceptionV3 and ConvNeXtTiny networks were trained with ImageNet~\cite{Imagenet} in advance, we reuse the trainable parameters from the first layer to the global pooling layer of these networks.
We then connect these pre-trained layers with a two dense layers as shown in the lower part in Figure~\ref{fig:f3} to perform the InceptionV3 and ConvNeXtTiny based classification models for the VSC task in this paper.
The InceptionV3 and ConvNeXtTiny based classifiers, which are finetuned on the downstream VSC task, are referred to as Vis-CONV and Vis-INC models, respectively.

Given the individual pre-trained models of Aud-MEL, Aud-GAM, Aud-CQT for audio input and Vis-CONV and Vis-INC for visual input, 
we remove header layers of these pre-trained models (i.e. The header layers of the pre-trained models are either the softmax layer or the final dense layer) to perform the Audio and Visual Backbones as shown in Figure~\ref{fig:f1}.
In the other words, when we feed an audio or visual data into the pre-trained models of Aud-MEL, Aud-GAM, Aud-CQT, Vis-CONV, Vis-INC, the feature maps extracted at the first fully connected FC(1024) or at second fully connected FC(10) are considered as the audio and visual embeddings as shown in Figure~\ref{fig:f1}.

\subsection{Phase II: Train the Fusion Layer and the Dense Layer}
\label{phase2}
In this Phase II, we aim to train the Fusion Layer and the Dense Layer as shown in the lower part of the Figure\ref{fig:f1}.
Regarding the Fusion Layer, it is used to combine audio and visual embeddings, which are extracted from the Audio and Visual Backbones, to generate an audio-visual embedding representing for one video input.
In this paper, we proposed three combination methods for the Fusion Layer.
Additionally, we have two types of audio/visual embeddings: The first type of audio/visual embeddings are extracted from the first fully connected layer FC(1024) of the pre-trained deep learning based models: Aud-MEL, Aud-GAM, Aud-CQT, Vis-CONV, Vis-INC; 
and the second type of audio/visual embeddings are extracted from the second fully connected layer FC(10) of these pre-trained deep learning based models.
As a result, we totally evaluate six types of Fusion Layer, referred to as $f_1, f_2, f_3, f_4, f_5$, and $f_6$.
While $f_1, f_2, f_3$ are three types of combinations for the first type of audio/visual embeddings, $f_4, f_5, f_5$ are for the second type of audio/visual embeddings.
Let consider $\{\mathbf{ae_g, ae_m, ae_c, ve_i, ve_c}\} \in R^{1024}$ as the the first type of audio and visual embeddings extracted from the the first fully connected layer FC(1024) of the Audio and Visual Backbones, the fusion functions of $f_1, f_2, f_3$ representing for the Fusion Layer are described by
\begin{small}
\begin{align}
    \label{eq:loss_func}
    f_{1} = \mathbf{ae_g}.\mathbf{w_1} + \mathbf{ae_m}.\mathbf{w_2} + \mathbf{ae_c}.\mathbf{w_3} + \mathbf{ve_i}.\mathbf{w_4} + \mathbf{ve_c}.\mathbf{w_5} + \mathbf{b},
\end{align}
\begin{equation}
\label{eq:loss_func}
    f_{2} = (\mathbf{ae_g}.\mathbf{w_1} + \mathbf{ae_m}.\mathbf{w_2} + \mathbf{ae_c}.\mathbf{w_3}).\mathbf{w_a} + \\ (\mathbf{ve_i}.\mathbf{w_4} + \mathbf{ve_c}.\mathbf{w_5}).\mathbf{w_v} + \mathbf{b},
\end{equation}
\begin{equation}
    \label{eq:loss_func}
    f_{3} = concat[(\mathbf{ae_g}.\mathbf{w_1} + \mathbf{ae_m}.\mathbf{w_2} + \mathbf{ae_c}.\mathbf{w_3}), \\ (\mathbf{ve_i}.\mathbf{w_4} + \mathbf{ve_c}.\mathbf{w_5})],
\end{equation}
\end{small}
where $\{\mathbf{w_1, w_2, w_3, w_4, w_5, w_a, w_v, b}\} \in R^{1024}$ are trainable parameters. 

Regarding the fusion function $f_1$, we assume that individual audio/visual embeddings have a linear relation across each dimension. 
Therefore, we apply the element-wise product between each trainable vector of $\mathbf{w_1, w_2, w_3, w_4, w_5}$ and each individual embedding before adding a bias $\mathbf{b}$.
By this way, a linear function, which helps to learn the relation of audio/visual embeddings across 1024 dimension, is established.
Meanwhile, in the fusion function $f_2$, we first apply the linear combination for only audio embeddings and for visual embeddings independently. 
Then, we again apply the linear combination for both audio and visual embeddings using trainable vectors of $\mathbf{w_a, w_v}$ and $\mathbf{b}$.
For the fusion function $f3$, we also first apply the linear combination for only audio embeddings and only visual embeddings independently.
We then concatenate two audio and visual embeddings to perform one audio-visual embedding.
The fusion functions $f_4, f_5, f_6$ share the same equation as $f_1, f_2, f_3$ respectively with the second type of audio/visual input embeddings of $\{\mathbf{ae_g, ae_m, ae_c, ve_i, ve_c}\} \in R^{10} $ and the trainable parameters of $\{\mathbf{w_1, w_2, w_3, w_4, w_5, w_a, w_v, b}\} \in R^{10}$.

The output of the Fusion Layer, known as the audio-visual embedding, is finally classified by a Dense Layer performed by a fully connected layer FC(10) and a Softmax layer as shown in the Figure~\ref{fig:f1}. 
Notably, as we freeze the Audio and Visual Backbones in the Phase II training process, the model is forced to learn the Fusion Layer and the Dense Layer.

\section{Experimental Results and Discussions}
\label{result}

\subsection{Implementation of deep learning models}
\label{setting}
We apply Tensorflow framework for implementing all deep learning based models in this paper.
As mixup~\cite{mixup2} data augmentation is used for audio spectrograms, image frames, and audio/visual embeddings to enforce classifiers, the labels of the augmented data are no longer one-hot.
We therefore use Kullback-Leibler (KL) divergence loss to train back-end classification models:
\begin{align}
    \label{eq:loss_func}
    LOSS_{KL}(\Theta) = \sum_{n=1}^{N}\mathbf{y}_{n}\log\left(\frac{\mathbf{y}_{n}}{\mathbf{\hat{y}}_{n}} \right)  +  \frac{\lambda}{2}||\Theta||_{2}^{2},
\end{align}
where $N$ is the training samples, $\Theta$ present the trainable network parameters, and $\lambda$ denotes the $\ell_2$-norm regularization coefficient. 
$\mathbf{y_{n}}$ and $\mathbf{\hat{y}_{n}}$  denote the ground-truth and the network output, respectively. 
All the training processes in this paper are run on two GeForce RTX 2080 Titan GPUs  using Adam method~\cite{Adam} for optimization.
\subsection{Datasets and evaluation metric}
\label{dataset}
This dataset is referred to as DCASE Task 1B Development, which was proposed for DCASE 2021 challenge~\cite{dcase_web}. 
The dataset is slightly imbalanced and contains both acoustic and visual information, recorded from 12 large European cities: Amsterdam, Barcelona, Helsinki, Lisbon, London, Lyon, Madrid, Milan, Prague, Paris, Stockholm, and Vienna. It consists of 10 scene classes: airport, shopping mall (indoor), metro station (underground), pedestrian street, public square, street (traffic), traveling by tram, bus and metro (underground), and urban park, which can be categorized into three meta-class of indoor, outdoor, and transportation.
The dataset was recorded by four recording devices simultaneously with the same setting of 48000 Hz sampling rate and 24 bit resolution. 
We obey the DCASE 2021 Task 1B challenge~\cite{dcase_web}, separate this dataset into training (Train.) subset for the training process and evaluation (Eval.) subset for the inference.
As regards the evaluation metric, the Accuracy (Acc.\%), which is commonly applied in classification tasks~\cite{dcase_web} and is also proposed for DCASE Task 1B challenge, is used to evaluate the AVSC task in this paper. 

\subsection{Experimental results and discussion}
\label{result}
\begin{figure*}[t]
    \centering
    \includegraphics[width =1.0\linewidth]{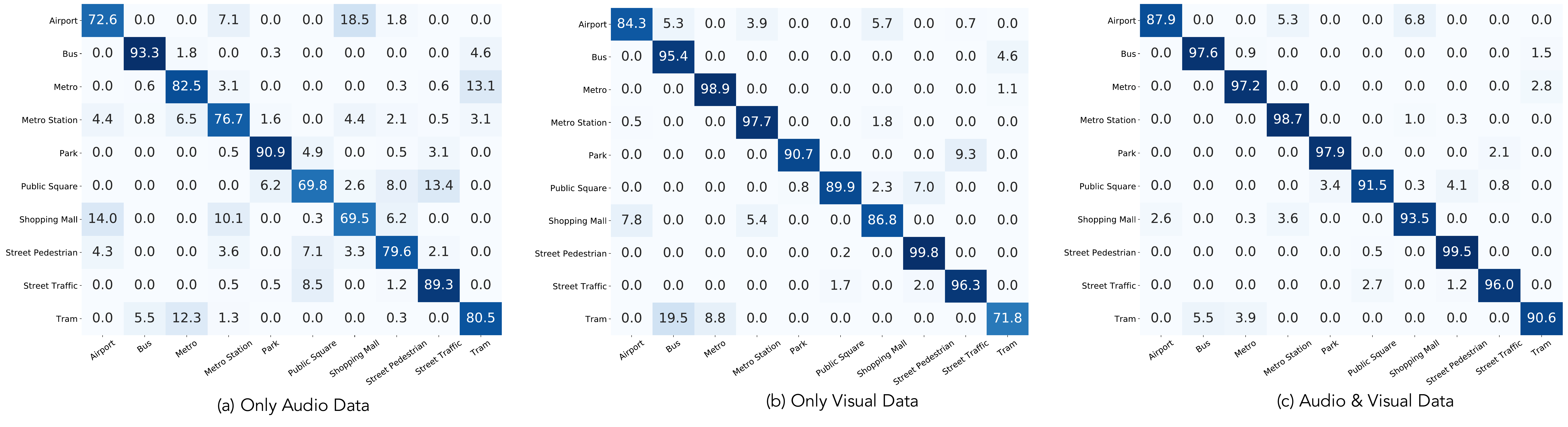}
    	\vspace{-0.5cm}
	\caption{The confusion matrix results of the propose systems using $f_4$ fusion method with only audio data (a), with only visual data (b), and with both audio and visual data (c)}
    \label{fig:t1}
\end{figure*}
We first evaluate the performance of our proposed systems with different types of fusion methods mentioned in Section~\ref{phase2}.
As the results show in Table~\ref{table:res1}, fusion methods of $f_4, f_5, f_6$ outperform $f_1, f_2, f_3$ respectively.
In the other words, the fusions of audio/visual embeddings extracted from the second fully connected layer FC(10) are more effective rather than the fusions of audio/visual embeddings from the first fully connected layer FC(1024).
We also see that the best accuracy score of 95.3\% is achieved from $f_4$ method which presents a linear combination of all five audio/visual embeddings. 

We then evaluate the performance comparison among the proposed systems using $f_4$ fusion of only audio data, of only visual data, of both audio and visual data.   
As the Figure~\ref{fig:t1} shows, the proposed AVSC system using only visual data (91.8\%) outperforms the system with only audio data (80.5\%) over almost categories, except of `Tram' and `Park'.
When both audio and visual data are used, this helps to improve the performance in all categories (Most categories record have accuracy more than 90\%, except ‘Airport’ with 88.0\%).

 \begin{table}[t]
	\caption{The performance of the proposed system (Acc.\%) with different types of fusion function: $f_1, f_2, f_3, f_4, f_5, f_6$.} 
	\vspace{-0.1cm}
	\centering
	\begin{tabular}{|l ||c |c |c ||c |c |c |} 
		\hline 
		\textbf{Category}     &\textbf{$f_1$}    &\textbf{$f_2$} &\textbf{$f_3$}  &\textbf{$f_4$}  &\textbf{$f_5$}  &\textbf{$f_6$}\\ 
		\hline 
		\hline 
		Airport 	          &87.9 &82.2  &85.4 &\textbf{87.9}   &86.8 &83.3         \\        
		Bus     	          &96.9 &96.3  &99.1 &97.6   &\textbf{99.1} &97.6          \\        
		Metro	 	          &97.5 &94.2  &96.9 &97.2   &\textbf{98.1} &96.9       \\        
		Metro Stattion 	      &98.2 &97.7  &98.2 &98.7   &\textbf{99.0} &97.9        \\        
		Park                  &95.6 &92.7  &93.8 &97.9   &\textbf{99.0} &98.2       \\        
		Public square         &92.0 &\textbf{91.7}  &89.7 &91.5   &87.9 &88.1       \\        
		Shopping mall 	      &91.0 &90.7  &88.9 &\textbf{93.5}   &92.5 &92.0      \\        
		Street pedestrian     &99.3 &\textbf{99.5}  &\textbf{99.5} &\textbf{99.5}   &\textbf{99.5} &99.0            \\        
		Street traffic 	      &96.3 &96.5  &\textbf{96.8} &96.0   &95.8 &95.5       \\        
		Tram 	              &83.1  &\textbf{91.2}  &84.1 &90.6   &90.3 &85.7         \\   
		\hline 
		\hline 
         Overall              &94.1  &93.7  &93.6 &\textbf{95.3}  &95.0 &93.9   \\
		\hline 
	\end{tabular}    
	\label{table:res1} 
\end{table}

We compare our best systems (i.e. using $f_4$ fusion) with the state-of-the-art systems.
As the Table~\ref{table:sota} shows, our proposed systems using only audio or using only visual data outperforms the state-of-the-art systems, records the accuracy of 80.5\% and 91.8\%, respectively.
Our proposed system using both audio and visual data achieves the top-2 after the system from~\cite{okazaki2021multi}.
However, the top-1 system~\cite{okazaki2021multi} presented an intensive ensemble of nine large deep learning models (EfficientNet, ResNeSt, and RegNet for directly training audio data; ResNet-6.4F, FesNetSt-50d, HRNet-W18 for directly training visual data; CLIP based networks of ResNet-101, ResNet-50x4 ViT-B32 for extracting visual embeddings), which requires nine individual processes as well as a post processing method for an inference.
Meanwhile, our proposed system combines five lighter models (3 residual-inception based models for audio data (36 M trainable parameters), InceptionV3 and ConvTiny based models for visual data (69.4 M trainable parameters)) and presents an end-to-end inference process.   

\section{Conclusion}
\label{conclusion}
We have proposed a deep learning based multimodal system with the two-phase training strategy for classifying daily life videos. 
Our proposed model, which makes use of a multi-spectrogram approach for audio data (i.e. MEL, GAM, and CQT) and multiple networks for visual data (InceptionV3 and ConvNeXtTiny), achieves the best performance of 95.3\% on the benchmark dataset of DCASE 2021 Task 1B. 
The experimental results prove that our proposed AVSC system is very competitive to the state-of-the-art systems and potential for applying to real-life applications.
\begin{table}[t]
    \caption{Compare our proposed AVSC system to the state-of-the-art systems (Acc.\%) with only using audio data, with only using visual data, and with using both audio \& visual data} 
        	\vspace{-0.2cm}
    \centering
    \scalebox{0.95}{
    \begin{tabular}{|l| c| c| c|} 
        \hline 
        \textbf{Authors}               & \textbf{Audio} & \textbf{Visual} & \textbf{Audio \& Visual}   \\ 
		\hline 
		\hline 
     	DCASE baseline \cite{ds_2021_1b} &65.1  &77.0  &77.4\\
     	Javier~\cite{naranjo2021squeeze} &69.0  &87.0  &90.0\\
		Zhou et al.~\cite{pp_av_03}      &-     &89.5  &91.6\\
		Hou et al.~\cite{pp_av_04}       &73.6  &88.9  &94.1\\
     	Wang et al.~\cite{pp_av_01}      &75.2  &80.3  &94.2\\
		Chen et al.~\cite{pp_av_02}      &78.0  &90.9  &94.6\\	
    	Soichiro~\cite{okazaki2021multi} &78.1  &90.9  &\textbf{96.1}\\	
     	\hline 
		\hline 
        Our best system     &\textbf{80.5} &\textbf{91.8}  & 95.3 \\        
		\hline 

    \end{tabular}
    }
    \label{table:sota} 
\end{table}


\bibliographystyle{IEEEbib}
\bibliography{refs}

\begin{thebibliography}{10}

\bibitem{cite_06}
Shiqing Zhang, Shiliang Zhang, Tiejun Huang, W.~Gao, and Qi~Tian,
\newblock ``Learning affective features with a hybrid deep model for
  audio–visual emotion recognition,''
\newblock {\em IEEE Transactions on Circuits and Systems for Video Technology},
  vol. 28, pp. 3030--3043, 2018.

\bibitem{cite_02}
Joon~Son Chung, A.~Senior, Oriol Vinyals, and Andrew Zisserman,
\newblock ``Lip reading sentences in the wild,''
\newblock in {\em IEEE Conference on Computer Vision and Pattern Recognition
  (CVPR)}, 2017, pp. 3444--3453.

\bibitem{cite_10}
Fabian~Caba Heilbron, Victor Escorcia, Bernard Ghanem, and Juan~Carlos Niebles,
\newblock ``Activitynet: A large-scale video benchmark for human activity
  understanding,''
\newblock in {\em IEEE Conference on Computer Vision and Pattern Recognition
  (CVPR)}, 2015, pp. 961--970.

\bibitem{sui2020real}
Jien-De Sui, Wei-Han Chen, Tzyy-Yuang Shiang, and Tian-Sheuan Chang,
\newblock ``Real-time wearable gait phase segmentation for running and
  walking,''
\newblock in {\em 2020 IEEE International Symposium on Circuits and Systems
  (ISCAS)}, 2020, pp. 1--5.

\bibitem{cite_04}
Naoya Takahashi, Michael Gygli, and L.~Van Gool,
\newblock ``Aenet: Learning deep audio features for video analysis,''
\newblock {\em IEEE Transactions on Multimedia}, vol. 20, pp. 513--524, 2018.

\bibitem{ds_2021_1b}
Shanshan Wang, Annamaria Mesaros, Toni Heittola, and Tuomas Virtanen,
\newblock ``A curated dataset of urban scenes for audio-visual scene
  analysis,''
\newblock in {\em ICASSP 2021 - 2021 IEEE International Conference on
  Acoustics, Speech and Signal Processing (ICASSP)}, 2021, pp. 626--630.

\bibitem{dcase_web}
Detection, Classification of~Acoustic~Scenes, and Events Community,
\newblock ``Dcase challenges task 1a,''
  \url{http://dcase.community/challenge2021}, 2021.

\bibitem{pp_av_01}
Qing Wang, Jun Du, Siyuan Zheng, Yunqing Li, Yajian Wang, Yuzhong Wu, Hu~Hu,
  Chao-Han~Huck Yang, Sabato~Marco Siniscalchi, Yannan Wang, and Chin-Hui Lee,
\newblock ``A study on joint modeling and data augmentation of multi-modalities
  for audio-visual scene classification,''
\newblock in {\em 2022 13th International Symposium on Chinese Spoken Language
  Processing (ISCSLP)}, 2022, pp. 453--457.

\bibitem{pp_av_02}
Chengxin Chen, Meng Wang, and Pengyuan Zhang,
\newblock ``Audio-visual scene classification using a transfer learning based
  joint optimization strategy,''
\newblock {\em arXiv preprint arXiv:2204.11420}, 2022.

\bibitem{pp_av_03}
Liguang Zhou, Yuhongze Zhou, Xiaonan Qi, Junjie Hu, Tin~Lun Lam, and Yangsheng
  Xu,
\newblock ``Attentional graph convolutional network for structure-aware
  audiovisual scene classification,''
\newblock {\em IEEE Transactions on Instrumentation and Measurement}, vol. 72,
  pp. 1--15, 2023.

\bibitem{pp_av_04}
Yuanbo Hou, Bo~Kang, and Dick Botteldooren,
\newblock ``Audio-visual scene classification via contrastive event-object
  alignment and semantic-based fusion,''
\newblock in {\em 2022 IEEE 24th International Workshop on Multimedia Signal
  Processing (MMSP)}, 2022, pp. 1--6.

\bibitem{lam12}
Lam Pham, Khoa Tran, Dat Ngo, Hieu Tang, Son Phan, and Alexander Schindler,
\newblock ``Wider or deeper neural network architecture for acoustic scene
  classification with mismatched recording devices,''
\newblock in {\em Proceedings of the 4th ACM International Conference on
  Multimedia in Asia}, 2022.

\bibitem{lam02}
Lam Pham, Huy Phan, Truc Nguyen, Ramaswamy Palaniappan, Alfred Mertins, and Ian
  Mcloughlin,
\newblock ``Robust acoustic scene classification using a multi-spectrogram
  encoder-decoder framework,''
\newblock {\em Digital Signal Processing}, vol. 110, pp. 102943, 2021.

\bibitem{mixup2}
Yuji Tokozume, Yoshitaka Ushiku, and Tatsuya Harada,
\newblock ``Learning from between-class examples for deep sound recognition,''
\newblock in {\em International Conference on Learning Representations (ICLR)},
  2018.

\bibitem{pham2022robust}
Lam Pham, Dusan Salovic, Anahid Jalali, Alexander Schindler, Khoa Tran, Canh
  Vu, and Phu~X Nguyen,
\newblock ``Robust, general, and low complexity acoustic scene classification
  systems and an effective visualization for presenting a sound scene
  context,''
\newblock {\em arXiv preprint arXiv:2210.08610}, 2022.

\bibitem{yu2023inceptionnext}
Weihao Yu, Pan Zhou, Shuicheng Yan, and Xinchao Wang,
\newblock ``Inceptionnext: When inception meets convnext,''
\newblock {\em arXiv preprint arXiv:2303.16900}, 2023.

\bibitem{keras_app}
Fran\c{c}ois Chollet et~al.,
\newblock ``Keras,'' \url{https://keras.io}, 2015.

\bibitem{Imagenet}
Olga Russakovsky, Jia Deng, Hao Su, Jonathan Krause, Sanjeev Satheesh, Sean Ma,
  Zhiheng Huang, Andrej Karpathy, Aditya Khosla, Michael Bernstein,
  Alexander~C. Berg, and Li~Fei-Fei,
\newblock ``{ImageNet Large Scale Visual Recognition Challenge},''
\newblock {\em International Journal of Computer Vision (IJCV)}, , no. 3, pp.
  211--252, 2015.

\bibitem{Adam}
Diederik~P. Kingma and Jimmy Ba,
\newblock ``Adam: A method for stochastic optimization,''
\newblock {\em CoRR}, vol. abs/1412.6980, 2015.

\bibitem{okazaki2021multi}
Soichiro Okazaki, Quan Kong, and Tomoaki Yoshinaga,
\newblock ``A multi-modal fusion approach for audio-visual scene classification
  enhanced by clip variants.,''
\newblock in {\em Proc. DCASE}, 2021, pp. 95--99.

\bibitem{naranjo2021squeeze}
Javier Naranjo-Alcazar, Sergi Perez-Castanos, Aaron Lopez-Garcia, Pedro
  Zuccarello, Maximo Cobos, and Francesc~J Ferri,
\newblock ``Squeeze-excitation convolutional recurrent neural networks for
  audio-visual scene classification,''
\newblock in {\em Proc. DCASE}, 2021, pp. 16--20.

\end{thebibliography}
\end{document}